\documentstyle[aps,multicol,epsf]{revtex}

\begin{document}
\title{Localization transitions in non-Hermitian quantum mechanics}
\author{Naomichi Hatano$^*$ and David R. Nelson}
\address{Lyman Laboratory of Physics, Harvard University,
Cambridge, Massachusetts 02138}
\date{\today}
\maketitle

\begin{abstract}
We study the localization transitions which arise in both one and two 
dimensions when quantum mechanical particles described by a random 
Schr\"odinger equation are subjected to a constant imaginary vector potential.
A path-integral formulation relates the transition to flux lines depinned
from columnar defects by a transverse magnetic field in superconductors.
The theory predicts that the transverse Meissner effect is accompanied by 
stretched exponential relaxation of the field into the bulk and a diverging
penetration depth at the transition.
\end{abstract}

\begin{multicols}{2}

Although forbidden in conventional quantum mechanics, exponentiated 
non-Hermitian quantum Hamiltonians do appear in the transfer matrices of 
classical statistical mechanics problems.
A nonequilibrium process can be described as the time-evolution of
a non-Hermitian system \cite{Kadanoff68}.
Another example is the $XXZ$ spin chain mapped onto the asymmetric
six-vertex model \cite{McCoy68}.

In this Letter, we investigate a non-Hermitian quantum Hamiltonian
with randomness.
The study is motivated by a mapping of flux lines in a $d+1$-dimensional 
superconductor to the world lines of $d$-dimensional bosons.
Columnar defects in the superconductor, which were introduced experimentally
in order to pin the flux lines \cite{Hardy91}, give rise to random potential 
in the boson system \cite{Nelson93a}.
Although the field component $\mbox{\boldmath{$H$}}_z$ parallel to the 
columns acts like a chemical potential for the bosons, the component 
perpendicular to the columns results in a constant imaginary vector potential
\cite{Nelson90}.

We study localization in this simple example of non-Hermitian 
quantum mechanics, and thereby show how a flux line is
depinned from columnar defects by an increasing perpendicular magnetic field 
$\mbox{\boldmath{$H$}}_\perp$.
It is generally believed that all eigenstates are localized in conventional
one- and two-dimensional non-interacting quantum systems with randomness.
On the other hand, it is almost obvious that a flux line is depinned from
defects by a strong perpendicular field component.
This indicates that the present non-Hermitian system 
has extended states in a large $\mbox{\boldmath{$H$}}_\perp$ region, and 
that there must be a delocalization transition at
a certain strength of $\mbox{\boldmath{$H$}}_\perp$.

The non-Hermitian Hamiltonian treated hereafter has the form
${\cal H}\equiv 
\left(\mbox{\boldmath{$p$}}+i\mbox{\boldmath{$h$}}\right)^2/(2m)
+V(\mbox{\boldmath{$x$}})$,
where $\mbox{\boldmath{$p$}}\equiv (\hbar/i)\mbox{\boldmath{$\nabla$}}$,
and $V(\mbox{\boldmath{$x$}})$ is a random potential.
The non-Hermitian field $\mbox{\boldmath{$h$}}$ originates in the 
transverse magnetic field as
$\mbox{\boldmath{$h$}}=\phi_0\mbox{\boldmath{$H$}}_\perp/(4\pi)$,
where $\phi_0$ is the flux quantum \cite{Nelson93a,Nelson90}.
Figure 1 shows a vortex whose ``world line'' is described by
this Hamiltonian with periodic boundary conditions in one dimension.
The mass $m$ is equivalent to the tilt modulus of the flux line.
The Planck parameter $\hbar$ corresponds to the temperature of the 
superconductor, while the inverse temperature of the quantum system 
corresponds to the thickness $L_\tau$ of the superconductor.
Interactions between many particles (or flux lines) can be treated 
approximately by forbidding multiple occupancy of localized state in a tight
binding model (see below) \cite{Nelson93a}.
Interactions in the delocalized regime will be discussed in a future
publication \cite{Hatano97}.

Since the field $\mbox{\boldmath{$h$}}$ acts like a vector potential, we 
define the current operator as $\mbox{\boldmath{$J$}}
\equiv -i\partial {\cal H}/\partial \mbox{\boldmath{$h$}}
=(\mbox{\boldmath{$p$}}+i\mbox{\boldmath{$h$}})/m$.
The imaginary part of the current describes the tilt slope of a flux line.
To see this, note first that
the position of the flux line $\mbox{\boldmath{$x$}}$
at the distance $\tau$ from the bottom surface
of the superconductor is given by
$\left\langle \mbox{\boldmath{$x$}} \right\rangle_\tau
\equiv {\cal Z}^{-1}
\langle \psi^f | e^{-(L_\tau-\tau){\cal H}/\hbar} 
\mbox{\boldmath{$x$}} e^{-\tau{\cal H}/\hbar} 
| \psi^i \rangle$,
where $\psi^i$ and $\psi^f$ describe boundary conditions at the bottom
and top surfaces ($\tau=0,L_\tau$) of the superconductor, 
respectively. The partition function is 
${\cal Z}\equiv \langle \psi^f | e^{-L_\tau{\cal H}/\hbar} | \psi^i \rangle$.
The commutation relation 
$[{\cal H},\mbox{\boldmath{$x$}}]=-i\hbar\mbox{\boldmath{$J$}}$
leads immediately to
$(\partial/\partial\tau) \left\langle \mbox{\boldmath{$x$}}\right\rangle_\tau
=-i\left\langle \mbox{\boldmath{$J$}} \right\rangle_\tau
={\rm Im}\,\left\langle \mbox{\boldmath{$J$}} \right\rangle_\tau$.

The total displacement of the flux line
between the bottom and the top surfaces is given by
$\left\langle \mbox{\boldmath{$x$}} \right\rangle_{L_\tau}
-\left\langle \mbox{\boldmath{$x$}} \right\rangle_0
=\hbar(\partial/\partial\mbox{\boldmath{$h$}}) \ln {\cal Z}
={\rm Im}\int_0^{L_\tau}\langle J \rangle_\tau d\tau$.
This quantity is an indicator of the delocalization transition;
the transverse displacement of a flux line 
must be order of the system size when it is depinned.

Let us first consider localized states in a small perpendicular field.
Assume the eigenfunctions $\psi_n(\mbox{\boldmath{$x$}})$ and the eigenvalues
$\varepsilon_n$ are known for $\mbox{\boldmath{$h$}}=0$.
For small $\mbox{\boldmath{$h$}}$ the right- and left-eigenvectors of 
${\cal H}$ are given by
$\psi^R_n(\mbox{\boldmath{$x$}};\mbox{\boldmath{$h$}})
=e^{\mbox{\boldmath{$h$}}\cdot\mbox{\boldmath{$x$}}/\hbar}
\psi_n(\mbox{\boldmath{$x$}};\mbox{\boldmath{$h$}}=0)$,
and
$\psi^L_n(\mbox{\boldmath{$x$}};\mbox{\boldmath{$h$}})
=e^{-\mbox{\boldmath{$h$}}\cdot\mbox{\boldmath{$x$}}/\hbar}
\psi^\ast_n(\mbox{\boldmath{$x$}};\mbox{\boldmath{$h$}}=0)$.
The energy eigenvalue $\varepsilon_n$ is unchanged under this 
``imaginary'' gauge transformation \cite{LeDousal}.
The imaginary gauge transformation applies even to many-body eigenvectors
of interacting systems \cite{Hatano97}.
However, the above wave functions $\psi^R_n$ and $\psi^L_n$ may diverge in 
$|\mbox{\boldmath{$x$}}|\to\infty$, and hence may not be normalizable.
The normalizablity condition is $|\mbox{\boldmath{$h$}}|<\hbar\kappa_n$,
where $\kappa_n$ is the inverse localization length of the state 
$\psi_n(\mbox{\boldmath{$x$}};\mbox{\boldmath{$h$}}=0)$.
Then the wave function (with the normalization 
$\int d^d\mbox{\boldmath{$x$}} \psi^R_n(\mbox{\boldmath{$x$}})
\psi^L_n(\mbox{\boldmath{$x$}})=1$) is approximately
\begin{equation}\label{gauge-20}
\psi^R_n(\mbox{\boldmath{$x$}})
\simeq 
\sqrt{\frac{(2\kappa_n)^d}{\Gamma(d)\Omega_d}}
e^{
\mbox{\boldmath{$h$}}\cdot
(\mbox{\boldmath{$x$}}-\mbox{\boldmath{$x$}}_n)/\hbar
-\kappa_n|\mbox{\boldmath{$x$}}-\mbox{\boldmath{$x$}}_n|},
\end{equation}
where $\mbox{\boldmath{$x$}}_n$ is the localization center 
for $\mbox{\boldmath{$h$}}=0$ and $\Omega_d$ is 
\begin{minipage}[t]{3.375in}
\epsfxsize=3.375in
\epsfbox{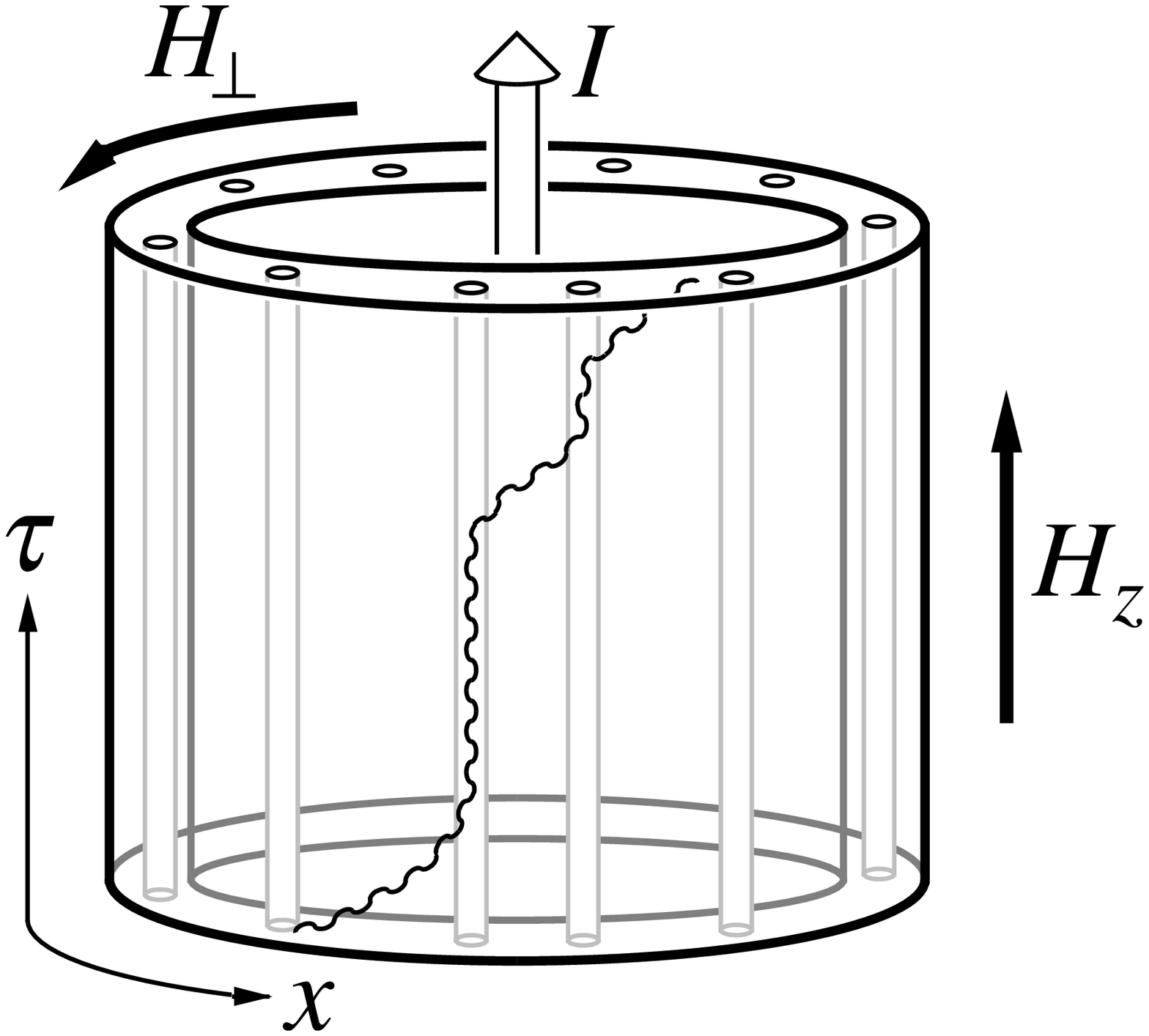}
\begin{small}
FIG.\ 1. 
One flux line (wavy curve) induced by the field 
$\mbox{\protect\boldmath{$H$}}_z$ and
interacting with columnar pins in a cylindrical superconducting shell with 
radial thickness smaller than the penetration depth of the defect-free
material.
The field $\mbox{\protect\boldmath{$H$}}_\perp$ is generated by the current 
$\mbox{\protect\boldmath{$I$}}$ threading the ring.
\end{small}
\vspace*{0.25in}
\end{minipage}
the total solid angle of the $d$-dimensional space.

We naturally regard the point $|\mbox{\boldmath{$h$}}|=\hbar\kappa_n$ as
the delocalization point of the state $\psi^R_n$.
In the region $|\mbox{\boldmath{$h$}}|\geq\hbar\kappa_n$, 
the eigenfunction becomes extended.
Hence we need to specify the boundary conditions in order to obtain
a well-defined wave function in the thermodynamic limit.

Let us consider imposing the periodic boundary condition
$\psi^R_n(L_x/2,y,\ldots)=\psi^R_n(-L_x/2,y,\ldots)$,
with the $x$ axis parallel to $\mbox{\boldmath{$h$}}$.
The one-dimensional periodic system is realized in the setup shown in
Fig.\ 1.
Except for the {\em imaginary} vector potential, this setup is the flux-line 
analogue of a mesoscopic metal ring threaded by a solenoid.
At the boundaries $x=\pm L_x/2$, the wave function of the form 
(\ref{gauge-20}) has a mismatch of the factor $e^{-(\kappa_n-h/\hbar)L_x}$.
In the region $h<\hbar\kappa_n$, this mismatch is exponentially small, 
and so is the change necessary to meet the periodic boundary condition.
In the region $h\geq\hbar\kappa_n$, however, 
the wave function changes drastically,
and a complex eigenvalue appears.
To understand this point, 
consider the case $|\mbox{\boldmath{$h$}}|\to\infty$, a limit in which
the random potential $V(\mbox{\boldmath{$x$}})$ may be neglected 
\cite{Hatano97}.
The periodic boundary condition is satisfied only by the extended function
$e^{i\mbox{\boldmath{$k$}}\cdot\mbox{\boldmath{$x$}}}$
with $k_\nu=2n_\nu\pi/L_\nu$, where $n_\nu$ is an integer and 
$L_\nu$ is the system size in the $x_\nu$ direction.
(If a right-eigenvector is
$e^{i\mbox{\boldmath{$k$}}\cdot\mbox{\boldmath{$x$}}}$, the corresponding 
left-eigenvector is $e^{-i\mbox{\boldmath{$k$}}\cdot\mbox{\boldmath{$x$}}}$.)
The eigenvalue is complex:
$\varepsilon(\mbox{\boldmath{$k$}})=
(\hbar\mbox{\boldmath{$k$}}+i\mbox{\boldmath{$h$}})^2/(2m)$.

The imaginary part of an eigenvalue has the following meaning.
The time evolution of an extended state is described as
$e^{i\mbox{\boldmath{$k$}}\cdot\mbox{\boldmath{$x$}}
-\tau \varepsilon/\hbar}$.
This must actually be a function of 
$\mbox{\boldmath{$k$}}\cdot(\mbox{\boldmath{$x$}}
-\tau\,{\rm Im}\,\langle\mbox{\boldmath{$J$}}\rangle)$, because
the corresponding flux line has the tilt slope 
${\rm Im}\,\langle\mbox{\boldmath{$J$}}\rangle$.
(The angular brackets here denote the expectation value 
with respect to the state with energy $\varepsilon$.)
We hence expect the relation
${\rm Im}\,\varepsilon\simeq
\hbar\mbox{\boldmath{$k$}}\cdot{\rm Im}\,\langle\mbox{\boldmath{$J$}}\rangle
\equiv-\hbar\mbox{\boldmath{$k$}}\cdot
(\partial/\partial \mbox{\boldmath{$h$}})\,{\rm Re}\,\varepsilon$.
Thus appearance of a complex eigenvalue is another indicator of the
delocalization transition.

It is instructive to solve the one-dimensional system with a single attractive
impurity:
${\cal H}\equiv -(\hbar\nabla-h)^2/(2m)-V_0\delta(x)$.
We first solve ${\cal H} \phi(x)=\varepsilon\phi(x)$ for $x\neq0$ to
obtain the two degenerate solutions $\phi_\pm(x)=e^{\pm ikx+hx/\hbar}$,
and then write the general solution as $\psi=A\phi_++B\phi_-$.
We impose the periodic boundary condition $\psi(L_x)=\psi(0)$
and the condition for the delta potential, 
$\psi'(L_x)-\psi'(0)=(2mV_0/\hbar^2)\psi(0)$.
A nontrivial solution arises whenever 
$k\left[\cosh(L_xh/\hbar)-\cos(L_xk)\right]+
(mV_0/\hbar^2)\sin(L_xk)=0$.
The ground state (defined by the lowest real part of the eigenvalue)
is localized only for $h<\hbar\kappa_{\rm gs}=mV_0/\hbar$.
All the other states are extended.
The localized ground-state energy is $\varepsilon_{\rm gs}=-mV_0^2/(2\hbar^2)$,
and the wave function takes the form 
$\psi^R_{\rm gs}(x)=\sqrt{\kappa_{\rm gs}}e^{-\kappa_{\rm gs}|x|+hx/\hbar}$ 
as $L_x\to\infty$.
The localization length in the direction of $\mbox{\boldmath{$h$}}$ grows as
$(\hbar\kappa_{\rm gs}-h)^{-1}$, and the state goes through a delocalization 
transition at $h=\hbar\kappa_{\rm gs}$. 
The ground state for $h>\hbar\kappa_{\rm gs}$
is extended with the energy $-h^2/(2m)$.
The corresponding flux line has the tilt slope $h/m$,
which is equal to the value in the impurity-free case.
The tilt slope of the ground state has a jump
at the delocalization point.

Extended states in the thermodynamic limit have a form close
to the pure case. The leading term is
$e^{ik_{n}x}/\sqrt{L_x}$ with $k_{n}=n\pi/L_x$, where
$n$ is odd for $h<\hbar\kappa_{\rm gs}$,
and is even for $h>\hbar\kappa_{\rm gs}$.
They also have the reflection term $e^{-ik_{n}x+2hx}$ in the region $x<0$.
The energy eigenvalue takes the same form as the pure system.
The above solution shows that a depinned flux line almost ignores 
the columnar defect.

Now we move on to the case of random potential.
For the purpose of numerical calculations, it is convenient to introduce
the non-Hermitian tight-binding model.
The second-quantized Hamiltonian is written in the form
\begin{eqnarray}\label{lattice-10}
{\cal H}&\equiv &-\frac{t}{2}
\sum_{\mbox{\boldmath{$x$}}}
\sum_{\nu=1}^d \left(
e^{\mbox{\boldmath{$h$}}\cdot\mbox{\boldmath{$e$}}_\nu/\hbar}
b^\dagger_{\mbox{\boldmath{$x$}}+\mbox{\boldmath{$e$}}_\nu}
b_{\mbox{\boldmath{$x$}}}
+e^{-\mbox{\boldmath{$h$}}\cdot\mbox{\boldmath{$e$}}_\nu/\hbar}
b^\dagger_{\mbox{\boldmath{$x$}}}
b_{\mbox{\boldmath{$x$}}+\mbox{\boldmath{$e$}}_\nu}
\right)
\nonumber\\
&+&\sum_{\mbox{\boldmath{$x$}}}
V_{\mbox{\boldmath{$x$}}} b^\dagger_{\mbox{\boldmath{$x$}}}
b_{\mbox{\boldmath{$x$}}},
\end{eqnarray}
where the $\{b^\dagger_{\mbox{\boldmath{$x$}}},b_{\mbox{\boldmath{$x$}}}\}$
are boson creation and annihilation operators \cite{boson}, 
and the $\{\mbox{\boldmath{$e$}}_\nu\}$
are the unit lattice vectors.
The hopping element is approximately given by \cite{Nelson93a}
$t\sim V_{\rm bind}\exp(-\sqrt{2mV_{\rm bind}}a/\hbar)$, 
where $V_{\rm bind}$ is 
the binding energy in the tight-binding approximation, and
$a$ is the lattice spacing.
We apply the periodic boundary conditions 
$b_{\mbox{\boldmath{$x$}}+N_\nu\mbox{\boldmath{$e$}}_\nu}=
b_{\mbox{\boldmath{$x$}}}$ for $\nu=1,2,\ldots,d$, where
$N_\nu\equiv L_\nu/a$.
The complex eigenvalues in this non-Hermitian system
always appear in conjugate pairs;
if there is a complex eigenvalue $\varepsilon$ with 
a right-eigenvector $\psi^R$, there is
also the eigenvalue $\varepsilon^\ast$ 
\begin{minipage}[t]{3.375in}
\epsfxsize=3.375in
\epsfbox{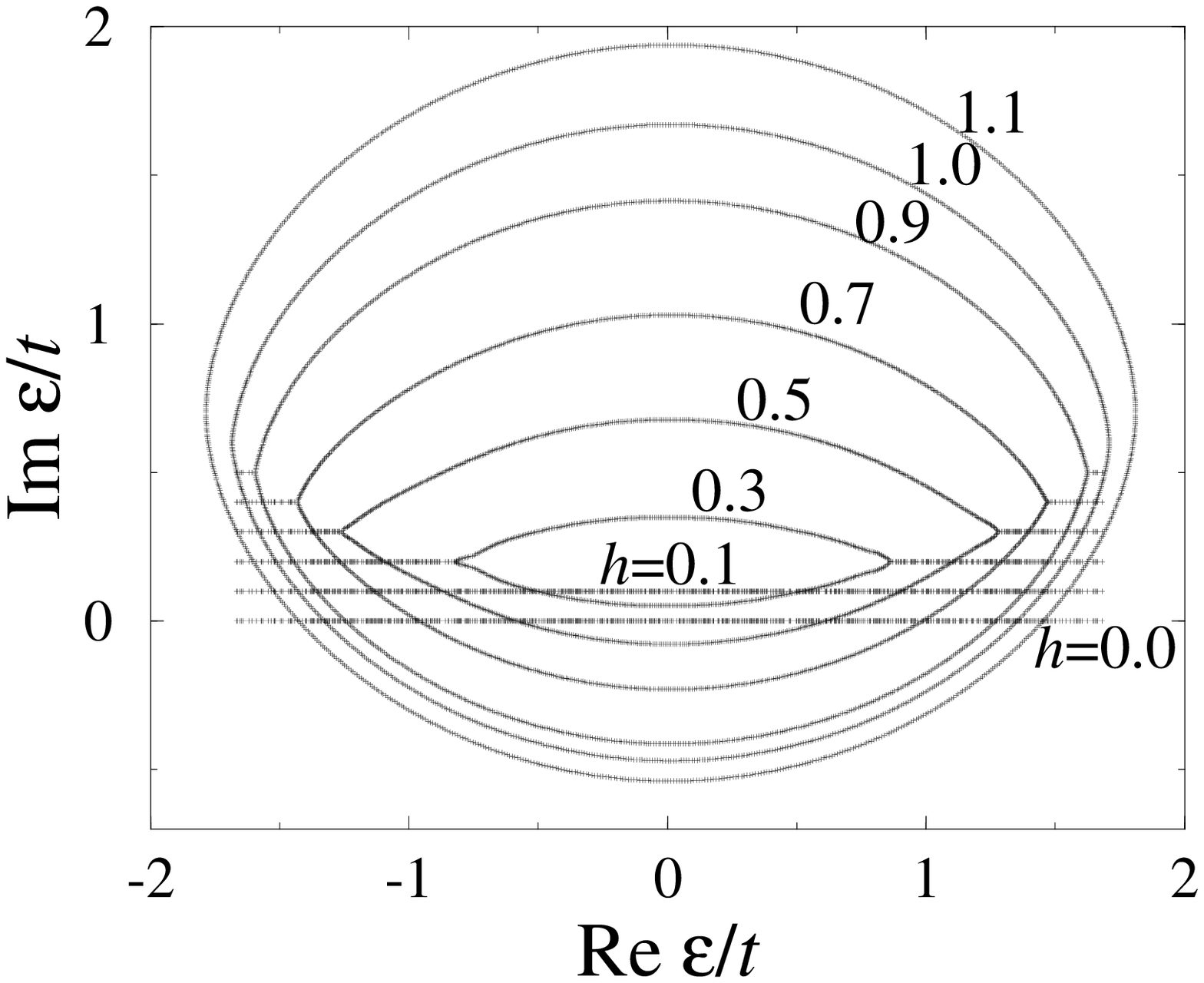}
\begin{center}
{\small (a)}
\end{center}
\epsfxsize=3.375in
\epsfbox{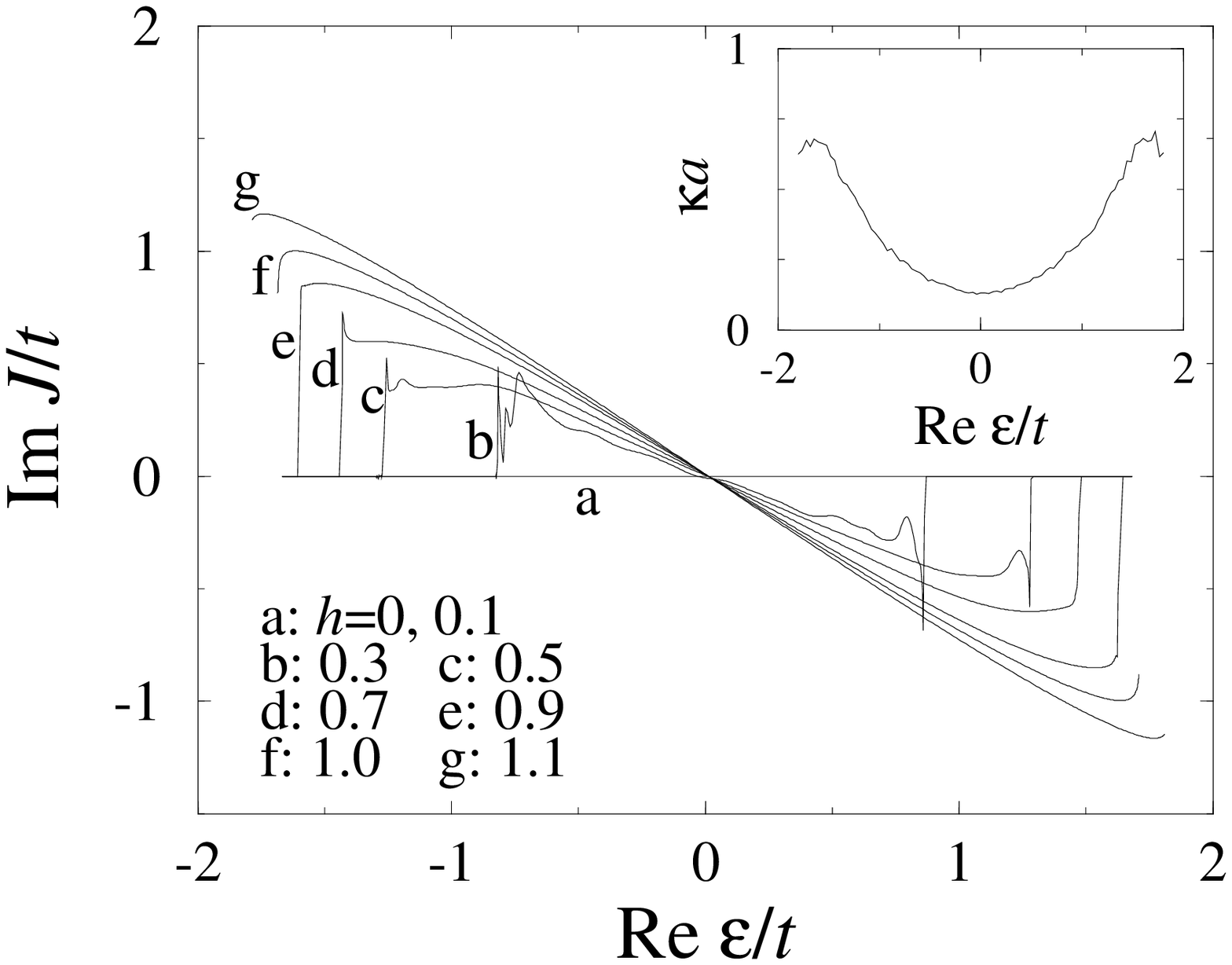}
\begin{center}
(b)
\end{center}
\begin{small}
FIG.\ 2. (a) ${\rm Im}\,\varepsilon$ vs.\ ${\rm Re}\,\varepsilon$ 
for a particular sample of the $d=1$ tight-binding model with 
$\Delta/t=1$ and $L_x=1000a$.
Each eigenstate is marked by a cross $+$.
Plots for different $\mbox{\protect\boldmath{$h$}}$ are offset for clarity.
(b) ${\rm Im}\,J$ (fluxon tilt) vs.\ ${\rm Re}\,\varepsilon$ 
for the same sample as in (a).
The inset shows $\kappa a$ vs.\ ${\rm Re}\,\varepsilon$ for 
$\Delta/t=1$, $L_x=500a$ and $\mbox{\protect\boldmath{$h$}}=0$, 
averaged over one hundred samples binned 
with energy window $0.04t$.
\end{small}
\vspace*{0.25in}
\end{minipage}
with the right-eigenvector
$(\psi^R)^\ast$.
This symmetry ensures that the partition function ${\cal Z}$ is real.
Another symmetry is 
${\cal H}(\mbox{\boldmath{$h$}})^T={\cal H}(-\mbox{\boldmath{$h$}})$.
Because of this symmetry, a right-eigenfunction of
${\cal H}(\mbox{\boldmath{$h$}})$ equals to the left-eigenfunction
of ${\cal H}(-\mbox{\boldmath{$h$}})$ with the same eigenvalue.
If $V_{\mbox{\boldmath{$x$}}}\equiv0$, the eigenvalues are
$\varepsilon=-t\sum_{\nu=1}^d\cos[(k_\nu+ih_\nu/\hbar)a]$
with $k_\nu=2n_\nu\pi/L_\nu$, where $n_\nu$ is an integer.
The eigenfunction takes the form
$\psi^R(\mbox{\boldmath{$x$}})\propto
\exp(i\mbox{\boldmath{$k$}}\cdot\mbox{\boldmath{$x$}})$.
Numerical calculations for $V_{\mbox{\boldmath{$x$}}}\neq0$
were carried out assuming that the random potential
$V_{\mbox{\boldmath{$x$}}}$ is uncorrelated in space,
and uniformly distributed in the range $[-\Delta,\Delta]$.
Results for random hopping models, with {\em no} site randomness,
are qualitatively similar \cite{Hatano97}.

Figure 2 (a) shows the eigenvalue spectrum for $d=1$.
The real eigenvalues indicate localized states,
while the complex eigenvalues indicate delocalized ones.
As discussed above,
the eigenvalues for the localized states are indeed independent of 
$\mbox{\boldmath{$h$}}$.
The behavior of the delocalized states, on the other hand, is similar
to impurity-free systems except near the mobility edges.
Figure 2 (b) shows the imaginary part of the current, 
or the tilt slope of the corresponding flux line
plotted against the real part of the energy.
The tilt slope has a jump at the mobility edges as in the one-impurity case.
The inset to Fig.\ 2 (b) shows the inverse localization length
for this problem when $\mbox{\boldmath{$h$}}=0$.
According to the delocalization criterion $h_c=\hbar\kappa$, a region of 
delocalized states should appear as $\mbox{\boldmath{$h$}}$ increases first 
at the band center and then move outwards, as we observe.

In two dimensions we again find bands of localized energies bounded by a 
mobility edge, although extended and localized states are apparently 
{\em mixed} in a complicated way near the band center.
Extended states again almost ignore impurities.
The case of an attractive impurity shows that the localized ground 
state exists in a certain region of small $\mbox{\boldmath{$h$}}$, 
and that the energy spectrum 
of the extended states is close to the impurity-free case.

An interesting question in the two-dimensional random case is whether a 
delocalized state is extended in both the $x$ and $y$ directions, 
or extended only in the direction of $\mbox{\boldmath{$h$}}$.
We conclude on the basis of the following argument that the former is the case.
If the field $\mbox{\boldmath{$h$}}$ is parallel to the $x$ axis
and a state is extended in this direction,
we may assume that the wave function has the approximate
form $L_x^{-1/2}e^{ik_xx}\phi(y)$ 
in order to accommodate periodic boundary conditions. 
This is indeed an eigenfunction of the partial Hamiltonian which has only 
the kinetic term in the $x$ direction.
We then calculate the effective Hamiltonian for $\phi(y)$ in degenerate 
perturbation theory, taking the kinetic term in the $y$ direction
and the random potential term as the perturbation.
We find that the effective Hamiltonian for $\phi(y)$ has a random potential 
term of the form $L_x^{-1}\int V(x,y)dx$.
The width of the probability distribution of this effective random potential
vanishes as $L_x^{-1/2}$ in the limit $L_x\to\infty$.
The flux line is a random walker in the $y$ direction, and hence $\phi(y)$ is 
an extended state.
Delocalization in directions both parallel and perpendicular to 
$\mbox{\boldmath{$h$}}$ is more readily observed numerically for 
$\mbox{\boldmath{$h$}}$
along the {\em diagonal} of a square lattice than for $\mbox{\boldmath{$h$}}$
parallel to $\mbox{\boldmath{$e$}}_x$ or $\mbox{\boldmath{$e$}}_y$
\cite{Hatano97}.

Spectra such as those in Fig.\ 2 (a) at intermediate values of 
$\mbox{\boldmath{$h$}}$ are relevant at low but {\em finite} concentrations 
of interacting flux lines, provided that we fill up the localized states in 
order of increasing energy up to energy $\varepsilon=\mu$, thus forbidding 
multiple occupancy of a single localized state \cite{Nelson93a}.
The system is then characterized by an average chemical potential 
$\mu=\mu(H_z)$ which
controls the flux-line density and separates occupied from unoccupied levels.
Once this chemical potential exceeds the mobility edge, however, we expect
spiral trajectories and 
``bose condensation'' of delocalized flux lines into the lowest-energy
extended state; more sophisticated methods of dealing with interactions
are then required \cite{Hatano97,Hwa93}.

Consider the probability distribution of a flux line near a free surface. 
At the distance $\tau$ from the bottom surface of the superconductor this 
probability may be written 
$P(\mbox{\boldmath{$x$}};\tau)
\equiv{\cal Z}^{-1}\langle\psi^f|e^{-(L_\tau-\tau){\cal H}/\hbar}
|\mbox{\boldmath{$x$}}\rangle
\langle \mbox{\boldmath{$x$}}|e^{-\tau{\cal H}/\hbar}|\psi^i\rangle$,
where we assume free boundary conditions at the bottom and the top surfaces:
$|\psi^i\rangle=|\psi^f\rangle
=\int d^d\mbox{\boldmath{$x$}}'|\mbox{\boldmath{$x$}}'\rangle$.
In the limit $L_\tau\to\infty$, 
the probability distribution of the most weakly bound flux line at the top 
surface is proportional to the right-eigenvector with the eigenvalue 
$\varepsilon=\mu$,
$P(\mbox{\boldmath{$x$}};L_\tau)
\propto\langle \mbox{\boldmath{$x$}}|\psi_\mu\rangle
=\psi^R_\mu(\mbox{\boldmath{$x$}})$,
while the distribution at the bottom surface is proportional to the 
left-eigenvector of the same state,
$P(\mbox{\boldmath{$x$}};0)
\propto\langle\psi_\mu|\mbox{\boldmath{$x$}}\rangle
=\psi^L_\mu(\mbox{\boldmath{$x$}})$.
The distribution far from these surfaces is given by 
$P(\mbox{\boldmath{$x$}};L_\tau/2)
=\psi^R_\mu(\mbox{\boldmath{$x$}})
\psi^L_\mu(\mbox{\boldmath{$x$}})$.
In the localized regime, the imaginary gauge transformation for
$\psi_\mu^R$ and $\psi_\mu^L$
describes the displacement of the flux 
line at the surfaces, while the bulk distribution 
$P(\mbox{\boldmath{$x$}};L_\tau/2)$ is 
{\em independent of} $\mbox{\boldmath{$h$}}$,
consistent with the transverse Meissner effect \cite{Nelson93a}.
These statements are approximate because we have neglected excitations of
occupied states. They are exact for the ground state.
The displacement of the flux lines near the top and bottom surfaces allows 
partial penetration of the transverse component of the magnetic field,
characterized by a transverse London penetration depth $\tau^\ast$.

In the case of one columnar defect (or a single impurity in a tight binding
model), the displacement of a flux line takes the form 
$(\langle\mbox{\boldmath{$x$}}\rangle_\tau
-\langle\mbox{\boldmath{$x$}}\rangle_\infty)
\cdot\mbox{\boldmath{$h$}}\sim
e^{-\tau\Delta\varepsilon_1/\hbar}$,
where $\Delta\varepsilon_1$ is the energy gap between the localized
ground state and the extended first excited state.
In the random case with finite concentration of flux lines, we consider the 
relaxation near a free surface $\tau=0$ of the most weakly bound line
to the highest-energy occupied state.
We show below that the displacement has the stretched exponential form
\begin{equation}\label{stretched}
(\langle\mbox{\boldmath{$x$}}\rangle_\tau
-\langle\mbox{\boldmath{$x$}}\rangle_\infty)
\cdot\mbox{\boldmath{$h$}}
\mathop{\sim}_{\tau\to\infty}
\exp[-a\left(\tau/\tau^\ast\right)^{1/(d+1)}],
\end{equation}
where $\langle\mbox{\boldmath{$x$}}\rangle_\infty$ is the center of the 
localized state in the bulk, the penetration depth is given by 
$\tau^\ast\equiv g(\mu)\hbar^{d+1}/(\hbar\kappa(\mu)-h)^d$, 
and $a$ is a constant of order of unity.
Here $g(\mu)$ and $\kappa(\mu)$ are respectively the density of states 
and the inverse localization length at the chemical potential.
Note that $\tau^\ast\sim\xi_\perp^d$, where $\xi_\perp\sim1/(\hbar\kappa-h)$
is the diverging surface localization length near the transition.

For the derivation of the above, we expand $P(\mbox{\boldmath{$x$}};\tau)$
with respect to the eigenfunctions.
In the limit $L_\tau\to\infty$, we have
$P(\mbox{\boldmath{$x$}};\tau)
\simeq \langle\psi_\mu|\mbox{\boldmath{$x$}}\rangle
\sum'_n c_n\langle\mbox{\boldmath{$x$}}|\psi_n\rangle
e^{-\tau\Delta\varepsilon_n/\hbar}$,
where $\Delta\varepsilon_n\equiv\varepsilon_n-\mu$, 
$\langle\psi_\mu|\mbox{\boldmath{$x$}}\rangle
=\psi_\mu^L(\mbox{\boldmath{$x$}})$ is the left-eigenvector describing the
shifted probability distribution of the surface, and the
$\{c_n\}$ are certain constants coming from the normalization factor.
The summation $\sum'_n$ is restricted to states with energies 
$\varepsilon>\mu$.
The asymptotic form (\ref{gauge-20})
for localized states now leads to 
$\left|\langle\mbox{\boldmath{$x$}}\rangle_\tau
-\langle\mbox{\boldmath{$x$}}\rangle_\infty\right|\sim
\sum'_n \tilde{c}_n e^{-f_n/\hbar}$, where 
$f_n(\tau)\equiv(\hbar\kappa_n-h\cos\theta_n)r_n+\tau\Delta\varepsilon_n$,
$r_n\equiv|\mbox{\boldmath{$x$}}_n-\mbox{\boldmath{$x$}}_\mu|$,
$\cos\theta_n\equiv\mbox{\boldmath{$h$}}\cdot
(\mbox{\boldmath{$x$}}_\mu-\mbox{\boldmath{$x$}}_n)
/|\mbox{\boldmath{$h$}}|r_n$,
and the $\{\tilde{c}_n\}$ are additional constant factors.
Upon setting $g(\mu)r_n^d\Delta\varepsilon_n\sim1$ \cite{Nelson93a},
we find that the two terms in $f_n$ compete with each other;
in other words, the further the flux line hops, the more the kink energy costs,
but the lower the binding energy of the state $n$ is.
Minimizing $f_n$ with respect to $r_n$ and $\theta_n$, we arrive at
Eq.\ (\ref{stretched}).
We expect that a similar relaxation process governs lines relaxing from the 
surface to occupied states with $\varepsilon<\mu$ as $\tau\to\infty$.
For $d=1$ and randomly distributed column {\em positions}, rare events like
exceptionally small hopping matrix elements will alter Eq.\ (\ref{stretched})
for very large $\tau$,
similar to flux lines pinned by parallel, randomly placed grain 
boundaries \cite{Marchetti95}.

The behavior of the flux-line tilt modulus (inverse boson superfluid density)
and flux-flow resistivity {\em above} the mobility edge may be obtained
\cite{Hatano97} using the methods of Ref.\ \cite{Hwa93}.

It is a pleasure to acknowledge discussions with B.I. Halperin, T. Hwa, 
and N. Schnerb.
One of us (D.R.N.) acknowledges interactions with P. Le Doussal and A. Stern
during the early stages of this research.
This work was supported by National Science Foundation, in part by the
MRSEC program through Grant DMR-9400396 and through Grant DMR-9417047.
N.H.\ acknowledges financial support from the Nishina Foundation.

\vskip 0.25in
$^*$On leave of absence from Department of Physics, University of Tokyo, 
Hongo, Bunkyo, Tokyo 113, Japan.

\end{multicols}

\begin{references}
\bibitem{boson}
Flux lines behave like bosons in the delocalized regime;
statistics are irrelevant when the lines are strongly localized
\protect\cite{Nelson93a}.
\end{references}
\end{document}